\documentstyle[11pt]{article}
\def\thefootnote{\fnsymbol{footnote}}

\def\thefootnote{\fnsymbol{footnote}}

\newcommand{\eq}{\begin{equation}}
\newcommand{\en}{\end{equation}}
\newcommand{\eqa}{\begin{eqnarray}}
\newcommand{\ena}{\end{eqnarray}}

\newcommand{\NP}[1]{Nucl.\ Phys.\ {\bf #1}}
\newcommand{\PL}[1]{Phys.\ Lett.\ {\bf #1}}

\newcommand{\PR}[1]{Phys.\ Rev.\ {\bf #1}}
\newcommand{\PRL}[1]{Phys.\ Rev.\ Lett.\ {\bf #1}}

\begin{document}

\hskip 11.5cm \vbox{\hbox{DFTT 27/94}\hbox{June 1994}}
\vskip 0.4cm
\centerline{\Large\bf Fluid Interfaces in the 3D Ising Model}
\centerline{\Large\bf as a Dilute Gas of Handles} 
\vskip 1.3cm
\centerline{ M. Caselle$^1$, F. Gliozzi$^1$ and U. Magnea$^{1,2}$
\footnote{e-mail: caselle, gliozzi, u\_magnea@to.infn.it}}
\vskip .6cm
\centerline{\sl  $^1$Dipartimento di Fisica 
Teorica dell'Universit\`a di Torino}
\centerline{\sl Istituto Nazionale di Fisica Nucleare, Sezione di Torino}
\centerline{\sl Via P. Giuria 1, I-10125 Turin, Italy}
\vskip 4mm
\centerline{\sl $^2$Department of Physics}
\centerline{\sl State University of New York at Stony Brook}
\centerline{\sl Stony Brook, New York 11794}
\vskip 2.5cm

\begin{abstract}
We study the topology of  fluid interfaces in the 3D Ising model in the 
rough phase. It turns out that such interfaces are 
accurately described as dilute 
gases of microscopic handles, and the stiffness of the interface
increases with the genus.
The number of configurations of genus $g$ 
follows a Poisson-like distribution. 
The probability per unit area for creating a handle is  well fitted in 
a wide range of the inverse temperature 
$\beta$ near the roughening point by an exponentially 
decreasing function of $\beta$. The procedure of summing over all 
topologies results in an effective interface whose squared width scales 
logarithmically with the lattice size. 
\end{abstract}
\noindent
\vfill
\eject

\newpage

\setcounter{footnote}{0}
\def\thefootnote{\arabic{footnote}}

\section{Introduction}
The properties of interfaces in 3D statistical systems have been a 
long-standing subject of research. 
One of the main reasons for this continuous 
interest probably lies in the fact that  
the physics of fluid interfaces can be very accurately 
described by field theoretic methods. Moreover, since interfaces are 
essentially two-dimensional objects, one is mainly interested in 2D
quantum field theories (QFT), a context in which significant 
improvements and new understanding have been achieved 
during the last few years (see, for instance, ref.~\cite{lh} for a 
comprehensive review). In particular it is by now generally accepted 
that the infrared (long range) properties of fluid interfaces are well 
described by a 2D, massless, free bosonic field theory. The bosonic 
field $h(x_1,x_2)$ is usually associated to
 the displacement of the interface from 
the equilibrium position as a function of the longitudinal coordinates
$x_1$ and $x_2$. This description can be obtained following 
a so called Solid on Solid (SOS) approximation of the interface or, 
equivalently,  
within a capillary wave approach. The crucial assumption in both 
these models is that the field $h$ must be a single-valued 
function of the longitudinal coordinates $x_1,x_2$ or, in other words, that 
the interface must have no overhangs and no handles.

Among the various realizations of fluid 
interfaces, a prominent role has been played in these last years by the 
3D Ising model. The reasons for this are first that the Ising model is  
in the same universality class 
as many physical systems, ranging from binary 
mixtures to amphiphilic membranes~\cite{book},
and second, that the Ising model, due to 
its intrinsic simplicity, allows fast and high statistics Montecarlo
simulations, so that very precise and discriminating comparisons can be 
made between theoretical predictions and numerical results.

Following this line, during the last year some high precision tests 
of the 
capillary wave model and of the consequent free bosonic QFT 
description have 
been performed~\cite{hp,cgv}. In all the various measurements a complete 
agreement was found between theoretical predictions and numerical 
results. In particular let us mention the
 analysis performed by M. Hasenbusch and K. Pinn~\cite{hp}
 of the logarithmic growth of the interface width with the 
longitudinal size $L_s$ of the interface in the rough phase (see below for 
details and definitions). This is indeed the simplest and 
most stringent prediction of the free bosonic model, being immediately 
related to the infrared divergence of the propagator of a free bosonic 
field in two dimensions. Moreover it has an appealing 
interpretation in terms 
of the Mermin-Wagner-Coleman theorem, as it is the signature 
of the fact that a 
continuous symmetry (in this case the translational invariance in the 
transverse direction) cannot be spontaneously broken in a 
two dimensional quantum field theory~\cite{L81}.

The impressive agreement of field theoretic predictions and Monte Carlo 
simulations prompted us to test the above mentioned assumption of a 
single valued behaviour of the interface. Indeed it is rather easy  to 
study the genus (namely the number of handles) of the interface in 
the Ising model. This can be done by a straightforward application of 
the Euler relation (see below) in the body-centered cubic (BCC) 
lattice where the interface is exactly self-avoiding, but it 
is also possible in the case of the simple cubic lattice (see below, and 
in ref.~\cite{cgv2}) through a suitable set of rules to separate the
interfaces along the self-intersection lines. This last case is the most 
interesting one since most of the published numerical results were
obtained on simple 
cubic lattices and a direct comparison  is thus possible.

In ref.~\cite{cgv2} we tested the assumption of a single-valued
interface in the region near the critical 
point where  the interface is expected to be almost delocalized. 
Surprisingly enough, we found that the assumption was apparently
completely wrong, 
that the probability of finding interfaces without handles was almost 
zero, that near the critical temperature the mean number of handles was 
very high (even of the order of several hundred on some of the 
lattices studied), 
and moreover that it was almost exactly proportional to the mean area 
of the interface, thus suggesting the picture of an interface 
``dressed'' by an huge number of microscopic handles. 

The simplest way to reconciliate this picture 
with the (now even more surprising) effectiveness of 
the free bosonic model in describing the interface, 
is to identify the field $h$ not with the 
microscopic ``bare'' interface of the Ising model (where with this term 
we indicate the Peierls contours separating the two phases of the Ising 
model), but with the ``dressed'' interface, in which all the handles 
contribute to give an intrinsic, finite thickness to the interface.

In ref.~\cite{cgv2} we explored some of the consequences of this picture, 
but it remained to understand whether such a dressed interface was 
a feature peculiar to the critical region or depended on 
the fact that all the 
temperatures we studied were of the same order as, or higher, than
 the percolation threshold. For this reason, in this paper we 
perform a systematic analysis of the genus dependence of various 
physical quantities {\it in the whole rough phase}. We will show that 
in the whole rough phase
 handles are present and that, similarly to what was observed in 
ref.~\cite{cgv2}, they are  microscopic (see sect. 4.1 below).
Besides this underlying common behaviour let us also stress some
important differences between the present work and ref.~\cite{cgv2}.
A first important difference from  the critical regime studied in 
ref.~\cite{cgv2} is that at these 
lower temperatures the density of handles is much smaller. 
As a consequence, the properties of these handles 
and the way in which their presence affects the behaviour of the whole 
interface, is very well described within the approximation of an uncorrelated 
handle distribution (see sect. 4.2 below). Second, the interfaces 
that we study in the present work are still rather well localized, their 
mean transverse width is much smaller than the lattice size $L_s$ which 
fixes their longitudinal size. Hence they can be essentially considered
 as two-dimensional objects.
 On the contrary in~\cite{cgv2}, due to the vicinity to the 
critical point, the interfaces were almost completely delocalized, they 
filled the whole lattice, their typical width being much larger than the 
lattice sizes in the longitudinal directions. As such they could not be 
considered at all as two-dimensional objects. This is precisely 
indicated by the $L_s$ dependence of the mean area of the interface (see 
sect. 4.3 below) whose power law behaviour is almost exactly $A\sim 
L_s^2$ in the present case 
(two-dimensional surface) and was a definitely 
larger number ($A\sim L_s^{3.7}$) in~\cite{cgv2}.

In accordance with the above  picture, we will show 
that agreement with the field theoretic predictions (in particular with 
the logarithmic law discussed above) is found only by 
summing over all the genera (see sect. 4.4 below). 
Thus the picture of an effective interface emerges, where the
microscopic structure is irrelevant, and which is well described by 
a bosonic field theory.
An important physical consequence of this analysis is that in the rough 
phase, but far enough from the bulk critical point so that the bulk 
critical length is still small, an increase of the genus of the 
interface corresponds to an increase of its stiffness. 

Let us conclude this introduction by noticing, as a side remark, the 
relevance of our results for the physics of self-avoiding surfaces; 
indeed  the fluid interfaces \cite{cgv2} or, more generally, the 
connected boundaries of the spin clusters \cite{dhmmpw} in the Ising or 
percolation models are at present the only systems where surfaces of 
high genera can be generated and are accessible to numerical 
simulations. Some  similarity between these systems
and the behaviour 
of 2D quantum gravity coupled to matter with central charge $c>1$
has been observed~\cite{{msy},{at}}.

\section{Capillary wave model}

The starting point of the capillary wave approach is the assumption that 
the  long-wavelength, 
transverse fluctuations of the interface (i.e., the capillary waves),
are described by an effective Hamiltonian 
proportional to the change they produce in the  area of the interface
\eq
H/k_BT=\int_0^{L_s} dx_1 
\int_0^{L_s}dx_2~\sigma(\theta)~ \left[\sqrt{1+\left(\frac{\partial h}{\partial x_1}\right)^2
+\left(\frac{\partial h}{\partial x_2}\right)^2}-1\right]~~,
\label{cw1}
\en
where the field $h(x_1,x_2)$ describes the displacement of the 
interface from 
the equilibrium position as a function of the longitudinal coordinates 
$x_1$ and $x_2$, $L_s$ is the 
size of the lattice in the longitudinal direction,
 $\sigma(\theta)$ is the 
(reduced) interface tension and $k_B$ is the Boltzmann constant.
 We have explicitly taken into 
account the  dependence of $\sigma$ on the angle 
$\theta(x_1,x_2)$ which the interface forms with the crystallographic 
plane. More precisely,  $\theta$ is defined in 
terms of the field $h$ as\footnote{In principle one would expect two 
different angles $\theta_1$ and $\theta_2$ for the two directions 
$x_1$, $x_2$, 
but here and in the following we assume a complete symmetry 
between the two longitudinal directions. This is justified since the two 
directions are equivalent on the $L_s^2\times L_t$ lattice we use.}:
\eq
\theta=arctan\left(\frac{d~h}{d~x}\right)
\en

The $\theta$ dependence becomes irrelevant near the critical point, where 
the bulk correlation length is large and rotational invariance is 
restored, but it turns out to be rather important in the remaining part 
of the rough phase, where it survives in the thermodynamic limit. 
The important parameter in this phase is the 
{\it stiffness} $k$ defined as:
\eq
k=\sigma(0)+\frac{d^2\sigma}{d~\theta^2}\vert_{\theta=0}~~~.
\en
In the limit $\beta\to\beta_c$ the $\theta$ dependence is negligibile 
and $k\to\sigma(0)$.

The interface free-energy increment due to interface 
fluctuations is then 
given by

\eq
F_{cw}=-k_BT ~log~ tr~e^{-H/k_BT}~~,
\label{free}
\en
The Hamiltonian of eq.~(\ref{cw1}) is  too difficult to be handled 
exactly. 
However, a crucial observation is  that this theory can be expanded in 
the adimensional parameter $({k L_s^2})^{-1}$ and the 
leading order term is, as anticipated, the Gaussian model. 
Then we replace eq.~(\ref{cw1}) with the $k L_s^2
\to\infty$ limit $H \to H_G$: 

\eq
 H_G/k_BT=\frac{k}{2}
\int_0^{L_s} dx_1 \int_0^{L_s} dx_2 ~ \left[\left(\frac{\partial h}
{\partial x_1}\right)^2
+\left(\frac{\partial h}{\partial x_2}\right)^2\right]~~.
\label{cw2}
\en

One of the most interesting predictions of the Gaussian model concerns
the average surface width $W$, defined by
\eq
W^2=\frac1{2L_s^4}\int_0^{L_s} d^2x\int_0^{L_s} d^2x'\langle 
\left(h(x_1,x_2)-h(x'_1,x'_2)\right)^2\rangle~~~.
\en
As is well known,
in the gaussian limit the square of the interfacial width grows 
logarithmically as
\eq
W^2=\frac{\beta_{eff}}{2\pi}log(L_s)+C_0
\label{w2}
\en
where $C_0$ is a constant and the "effective temperature" $\beta_{eff}$ 
coincides in this limit with the inverese of the stiffness:
\eq
k=\frac{1}{\beta_{eff}}~~~.
\label{kb}
\en
According to the widely accepted conjecture that the interface 
models near the roughening transition belong to the universality class 
of Kosterlitz and Thouless, it is expected that at the roughening point 
$\beta_{eff}=\frac2\pi$.

\section{The simulation}

We studied the 3D Ising model defined by the partition function
\eq
Z=\sum_{s_i=\pm 1} e^{-\beta H}
\en
with the standard Hamiltonian, with nearest neighbour interaction only,
\eq
H=-\sum_{<i,j>}s_is_j~~~,
\en
using a cluster algorithm.
We chose a simple cubic geometry whose drawback is that 
it does not allow 
a simple evaluation of the genus of the interface, but which has the 
important advantage that several properties of the model are known with 
very high precision. In particular
the Ising model  is known to have a second order bulk phase transition at 
$\beta_c=0.221652(3)$ and a roughening transition at $\beta_r=0.4074(3)$ 
\cite{IM,hp}.

The simulations were performed on a $L_s^2\times L_t $ lattice, 
with $L_s$ $(L_t)$ the lattice spacing in the longitudinal (transverse) 
direction. 
We chose periodic boundary 
conditions in the longitudinal directions and ${\it antiperiodic}$ in the
transverse 
direction. This forces the formation of an odd number of interfaces in 
this direction (actually, in almost all our configurations only one 
interface was present).
We fixed $L_t=32$, which in the 
range of $\beta$ that we studied is always much larger than the mean 
interface width $W$. 
We did two  different sets of measurements. First we  
systematically studied 
the formation of interfaces in the rough phase, choosing 
a rather small value of the longitudinal lattice spacing: $L_s=14$, but 
keeping high statistics (20,000 measurements for each value of $\beta$)
 and a very fine resolution in $\beta$. In particular, in order to 
better describe the region near the critical point, we chose a
finer resolution
close to $\beta_c$. A simple 
way of doing this is to choose a 
constant resolution $\Delta\beta_{dual}=0.01$  
in the dual coupling constant  $\beta _{dual}$, defined by
\eq
\beta_{dual}=-\frac{1}{2}~log~tanh~\beta~~~.
\en
We studied the range $0.34\leq\beta_{dual}\leq 0.70$ which corresponds
to $0.5582\geq\beta\geq 0.2518$~. The precise
 correspondence between $\beta$ and 
$\beta_{dual}$, for all the simulations we did, can be found in 
Table~II.

Second, we selected two values of $\beta$, for which the interface
was studied for varying lattice sizes $L_s$ in the longitudinal 
direction. They are listed in Table~I, together with the 
values of $\beta_{eff}$ (which we shall discuss in the next section) 
obtained by interpolating the data reported in Table~XII of ref.~\cite{hp}.
\def\ul{\underline}
\vskip0.3cm
$$\vbox {\offinterlineskip
\halign  { \strut#& \vrule# \tabskip=.5cm plus1cm
& \hfil#\hfil 
& \vrule# & \hfil# \hfil &
& \vrule# & \hfil# \hfil &\vrule# \tabskip=0pt \cr \noalign {\hrule}
&& $\beta$ && $L_s$  && $\beta_{eff}$ 
& \cr \noalign {\hrule}
&& $0.3536$ && $\ul{14},\ul{16},\ul{18},\ul{20},\ul{22},
\ul{24},\ul{48}$  && $1.45$ 
& \cr \noalign {\hrule}
&& $0.3175$ && $\ul{14},16,18,\ul{20},22,\ul{24},
\ul{32},38,\ul{48}$  
&& $2.26$ & \cr \noalign {\hrule}
}}$$
\begin{center}
{\bf Table~I.} {\it Set of longitudinal lattice sizes $L_s$ for which 
interfaces were studied. In the first column the corresponding values of 
$\beta$ appears, and in the last column, 
the corresponding values of $\beta_{eff}$ (see text) obtained 
in~\cite{hp} are presented. For the underlined lattice sizes  also the 
interfacial width was measured.}
\end {center}
\vskip0.3cm

In each configuration we first 
isolated the interface. This can be easily done by reconstructing all the 
spin clusters of the configuration, keeping the largest  one and 
flipping the others. By a repeated application of this algorithm
 we finally eliminated all the bubbles in the configuration and 
obtained two pure phases, separated by one interface. We shall call the 
resulting configuration a ``cleaned'' one.  
Then in this cleaned 
configuration we 
measured the area $A$, the genus $g$ and 
the width $W$ of the interface. Let us briefly describe how these 
observables were measured.

\begin{description}

\item{$Area$}

The area is the simplest observable to measure. Its value is given
by the number of frustrated links of the cleaned configuration.

\item{$Genus$}

If the interface were a self-avoiding surface (and this is the case, 
for instance, on the BCC lattice) its genus $g$ would be 
simply given by the Euler relation\footnote{ Notice 
that, due to the periodic boundary conditions in the longitudinal 
directions, the interfaces cannot have the
 topology of a sphere. The simplest possible interface has the 
topology of a torus. This means that $g\geq 1$ and that $\chi(S)$ cannot 
be positive.}:
\eq
F~-~E~+~V~=~2~-~2g~\equiv~ \chi(S)
\en
where $F$ is the number of faces (hence frustrated links in the dual 
framework), $E$ the number of edges, $V$ the number of vertices, and 
$\chi(S)$ the Euler characteristic of the interface 
$S$.
On the simple cubic lattice that we are studying this is not the case, 
and in general the interface has many self-intersections. We have shown 
in~\cite{cgv2} how to deal with this problem and in particular we  
gave a local, consistent, prescription for reconstructing from  the 
self-intersecting interface an equivalent self-avoiding surface.
The term ``equivalent'' means that it has  the same area
and width. This equivalence can be stated in a rigorous, 
 geometrical way in the framework of the Dhen 
lemma, which states that for any given self-intersecting 
surface $S_1$ with a self-avoiding boundary and for any real, positive
$\epsilon$, 
there exist infinitely many self-avoiding surfaces $S_n$ whose distance 
from $S_1$ in the space of parameters is smaller than $\epsilon$. 
Let us stress that our procedure is not unique and that in general a 
self-intersecting surface can be mapped into different equivalent  
self-avoiding surfaces, with the same area and width, but with 
different genus. Indeed our procedure turns out to give  a consistent 
recipe for  cutting the surface along the self-intersection lines and 
separating the resulting parts. 
These self-intersection lines are of two types: those that divide the 
surface in two disconnected pieces (and these can be eliminated without 
ambiguities), and those that do not.
It can be shown that the latter
give rise to a variation of the genus $\Delta 
g$ which can take the values $\Delta g=-1,0,1$. 
This is not strange since the 
self-intersection lines can be thought of as separating degenerate handles 
from the surface. Depending on how one ``cuts open'' this 
self-intersection line to create a self-avoiding surface, these degenerate
handles may either become real handles, or they may be ``reabsorbed'',
thus creating a ``pocket'' on the surface. 
These two possible resulting surfaces are characterized by different 
genus. Our reconstruction algorithm chooses between the two according to 
the sign of the local magnetization (for details see ref.~\cite{cgv2}).  
Once  a procedure for determining the genus  is established
 one can count the 
number of configurations of given genus $N(g)$ in the sample. 
This observable will play a major role in the following analysis.

\item{$Width$}

The definition of the interfacial width  is not unique.
In order to make the comparison of results easy, we have chosen the 
same definition as the authors of ref.~\cite{hp}. 
Let us define the mean magnetization in each slice in the $t$ direction
 of the lattice:
\eq
M(t)=\frac{1}{L_s^2}\sum_{x,y=1}^{L_s}s(x,y,t)
\en

 Let us then 
introduce an auxiliary coordinate $z$ for the $t$ direction, 
that assumes 
half integer values and labels the positions between adjacent lattice 
layers perpendicular to the transverse direction. 
Let us choose the origin
$z=0$ to be in the middle of the lattice, 
namely between the slices located 
at $t=L_t/2$ and $t=L_t/2+1$. This means that in terms of $z$ the above 
described 
slices of the lattice are labelled by $z=-\frac{L_t+1}{2},
~-\frac{L_t-1}{2}\cdots -\frac{1}{2},\frac{1}{2}\cdots 
\frac{L_t-1}{2}$ .
 Let us then, following the procedure described in \cite{hp},  
shift the interface to the $z=0$ position, keeping
track of 
the antiperiodic boundary conditions. 
The above shift is is not necessary , but reduces the spread in the 
measured values of the interface thickness, thereby diminishing the 
noise.
We can define a normalized magnetization 
gradient as:
\eq
\rho(z)=\frac{M(z+1/2)-M(z-1/2)}{M(-L_t/2)-M(L_t/2)}
\en
In terms of this gradient the square of the interfacial width can be 
written as
\eq
W^2=\left\langle\sum_z\rho(z) z^2-(\sum_z\rho(z) z)^2\right\rangle~~~,
\en
where the normalization has been chosen such that $W^2$ matches 
with the microscopic definition (\ref{w2}).
\end{description}

All the quoted errors were obtained with a standard jack-knife procedure.

\section{Results }

\subsection{ $\beta$ dependence of mean area and genus}

In Figs. 1 and 2 we report our results on the $\beta$ dependence, at 
fixed value 
of $L_s$, of the genus profile and mean area of the interfaces. 
As expected, for large values of $\beta$ 
(hence small temperatures) the interface is strongly 
constrained to the immediate neighbourhood of the 
crystallographic plane; the genus is always
one (recall the periodic boundary conditions in the longitudinal 
directions), 
and the mean area is almost exactly that $(L_s^2=196)$ of   
the crystallographic plane. As $\beta$ decreases the probability of 
formation of handles on the interfaces in the sample increases and 
correspondingly also the mean area of the interface. This 
indicates that the  fluctuations of the interface are getting larger and 
larger. There are some interesting properties of the interface
 which can be observed by studying these results:
\begin{description}
 \item{a]}
The genus of the interface is $not$ an order parameter of the 
roughening transition. Also for $\beta>\beta_r$ we find interfaces with 
genus greater than 1 and their number grows as $L_s$ increases.

\item{b]}
For any given value of $\beta$ we see that the mean area of the 
interface increases as the genus increases. Looking at Fig.~3 
one can see 
that the mean area differences between surfaces of genus $g$ and $g+1$:
$\Delta A\equiv A(g+1)-A(g)$ are essentially constant as a function 
of $g$, and grow slowly with $\beta$. This suggests to use these 
differences as an indicator of the mean size of a typical handle, which 
turns out to range from $\Delta A \sim 15 - 20$ 
near the roughening point, to
$\Delta A \sim 25-30$ in the region around $\beta=0.26$.
 Hence the typical handle 
is microscopic with respect to the size of the interface. 
Moreover, the fact that $\Delta A$ is almost independent of $g$ 
suggests that the various microscopic handles can be considered as 
independent and non-interacting. Thus adding a new handle, simply 
amounts to adding the same number 
$\Delta A$ of plaquettes to the surface.
While the mean size of the handles increases as $\beta$ decreases, 
the ratio $\Delta A/A$
 remains almost constant in the whole rough  region $\beta<\beta_r$
(see Fig.~4), thus indicating that 
the increase of $\Delta A$ is simply related to the overall increase 
of the mean area: the handles are getting ``rough'' just 
like the whole interface. 
\end{description}

\subsection{ Genus distribution function}

The above results suggest that, if $p$ 
is the probability of having one handle, then $p^n/n!$ is the 
probability 
of having $n$ handles, and that this probability $p$ should be 
proportional to the area $A$ of the surface. This leads us to conjecture
 the following ``Poisson-like'' distribution for the number of 
configurations of given genus $N(g)$ in the sample :
\eq
N(g)=C~\frac{{(\mu(\beta) A(g))}^{g-1}}{(g-1)!}
\label{dist}
\en
where $C$ is a normalization constant which does not depend on the 
genus $g$, $\mu(\beta)$
 is the probability per plaquette of having a handle and, 
as mentioned above, due to the periodic boundary 
conditions the number of microscopic handles is $n=g-1$.

In the limit $\Delta A\to 0$, in which $A$ does not depend on 
$g$ eq.~(\ref{dist}) would exactly become a Poisson distribution:
\eq
N_P(g)=N_s\frac{{\lambda}^{g-1}e^{-\lambda}}{(g-1)!}
\label{Poisson}
\en
where $N_s$ is the size of the sample,
 $\lambda=\mu A$ and $C=N_s exp(-\lambda)$.

As can be seen by looking at Figs. 5 and 6 our data are in remarkable 
agreement with eq.~(\ref{dist}). Moreover, in the samples corresponding
to smaller 
$\beta$ or larger $L_s$ in which $\Delta A$ cannot be neglected,
one can easily see that eq.~(\ref{dist}) gives a much better agreement 
with the data than eq.~(\ref{Poisson}). A crucial test of 
eq.~(\ref{dist}), is to show that the parameter $\mu$ only depends on 
$\beta$ and that the dependence on the lattice size $L_s$ is completely 
taken into account by the area factor. This can be tested by looking at 
the data at $\beta=0.3175$. As $L_s$ increases from 14 to 48 the function 
$N(g)$ changes dramatically, but it is always well described by 
eq.~(\ref{dist}) with the same value of the parameter: $\mu\sim 0.0012$, 
(see Fig.~6)
\footnote{
Due to the lack of statistics for higher genera, we do not  quote 
errors for $\mu$ and $\l_h$. In any case we estimate that an upper bound 
for such errors should be of the order of 5\%.}.  
 The function $N(g)$ is 
thus completely determined by the parameter $\mu(\beta)$ whose values are 
listed in Table~II and plotted in Fig.~7. The simplest hypothesis 
which can be made on the  $\beta$ dependence of $\mu$ is that it should 
be of the form: $\mu\propto exp(-2\beta l_h)$, where $\exp(-2\beta)$ is 
the cost of frustrating one link (hence create a new plaquette in the
surface) and we define $l_h$ to be the mean size of a typical handle.
This means that for consistency we expect $l_h\sim \Delta A$.   
It is rather interesting 
to notice that, even if this hypothesis is very crude, it describes 
surprisingly well the actual behaviour of $\mu$, at least in the rough 
region. Indeed it can be seen by looking at Fig.~7 that there is a wide 
range of values of $\beta$, starting from the roughening point, up to 
$\beta\sim 0.28$ in which $\mu$ is well described by the law:
\eq
\mu(\beta)=\mu_0~e^{-2\beta~l_h}
\en
with $l_h\sim 14$ and $\mu_0 \sim 7.5$, 
and that, as expected, $\l_h\sim \Delta A$. Moreover, as the critical 
point is approached and as  $\Delta A$ increases, also 
$\l_h$ increases (see the smallest values of $\beta$ in Fig.~7).

$$\vbox {\offinterlineskip
\halign  { \strut#& \vrule# \tabskip=.5cm plus1cm
& \hfil#\hfil
& \vrule# & \hfil# \hfil &
& \vrule# & \hfil# \hfil &\vrule# \tabskip=0pt \cr \noalign {\hrule}
&& $\beta_{dual}$ && $\beta$  && $\mu\times 1000$   & \cr 
\noalign {\hrule}
   && $0.34$ && $0.5582$ && $$ & \cr \noalign {\hrule}
   && $0.35$ && $0.5448$ && $0.0002$ & \cr \noalign {\hrule}
   && $0.36$ && $0.5318$ && $0.0009$ & \cr \noalign {\hrule}
   && $0.37$ && $0.5192$ && $0.0013$ & \cr \noalign {\hrule}
   && $0.38$ && $0.5071$ && $0.0031$ & \cr \noalign {\hrule}
   && $0.39$ && $0.4953$ && $0.0058$ & \cr \noalign {\hrule}
   && $0.40$ && $0.4839$ && $0.0053$ & \cr \noalign {\hrule}
   && $0.41$ && $0.4728$ && $0.0086$ & \cr \noalign {\hrule}
   && $0.42$ && $0.4620$ && $0.0154$ & \cr \noalign {\hrule}
   && $0.43$ && $0.4515$ && $0.021$ & \cr \noalign {\hrule}
   && $0.44$ && $0.4414$ && $0.028$ & \cr \noalign {\hrule}
   && $0.45$ && $0.4315$ && $0.036$ & \cr \noalign {\hrule}
   && $0.46$ && $0.4219$ && $0.055$ & \cr \noalign {\hrule}
   && $0.47$ && $0.4125$ && $0.079$ & \cr \noalign {\hrule}
   && $0.48$ && $0.4034$ && $0.111$ & \cr \noalign {\hrule}
}}$$
\vskip-0.2cm
\centerline {\bf a}

$$\vbox {\offinterlineskip
\halign  { \strut#& \vrule# \tabskip=.5cm plus1cm
& \hfil#\hfil
& \vrule# & \hfil# \hfil &
& \vrule# & \hfil# \hfil &\vrule# \tabskip=0pt \cr \noalign {\hrule}
&& $\beta_{dual}$ && $\beta$  && $\mu\times 1000$   & \cr 
\noalign {\hrule}
   && $0.49$ && $0.3946$ && $0.139$ & \cr \noalign {\hrule}
   && $0.50$ && $0.3860$ && $0.182$ & \cr \noalign {\hrule}
   && $0.51$ && $0.3776$ && $0.23$ & \cr \noalign {\hrule}
   && $0.52$ && $0.3694$ && $0.29$ & \cr \noalign {\hrule}
   && $0.53$ && $0.3614$ && $0.35$ & \cr \noalign {\hrule}
   && $0.54$ && $0.3536$ && $0.45$ & \cr \noalign {\hrule}
   && $0.55$ && $0.3461$ && $0.55$ & \cr \noalign {\hrule}
   && $0.56$ && $0.3387$ && $0.67$ & \cr \noalign {\hrule}
   && $0.57$ && $0.3314$ && $0.84$ & \cr \noalign {\hrule}
   && $0.58$ && $0.3244$ && $1.00$ & \cr \noalign {\hrule}
   && $0.59$ && $0.3175$ && $1.20$ & \cr \noalign {\hrule}
   && $0.60$ && $0.3108$ && $1.42$ & \cr \noalign {\hrule}
   && $0.61$ && $0.3043$ && $1.71$ & \cr \noalign {\hrule}
   && $0.62$ && $0.2979$ && $2.02$ & \cr \noalign {\hrule}
   && $0.63$ && $0.2917$ && $2.36$ & \cr \noalign {\hrule}
   && $0.64$ && $0.2856$ && $2.8$ & \cr \noalign {\hrule}
   && $0.65$ && $0.2796$ && $3.3$ & \cr \noalign {\hrule}
   && $0.66$ && $0.2738$ && $3.8$ & \cr \noalign {\hrule}
   && $0.67$ && $0.2681$ && $4.2$ & \cr \noalign {\hrule}
   && $0.68$ && $0.2625$ && $4.6$ & \cr \noalign {\hrule}
   && $0.69$ && $0.2571$ && $5.1$ & \cr \noalign {\hrule}
   && $0.70$ && $0.2518$ && $6.0$ & \cr \noalign {\hrule}
}}$$
\vskip-0.2cm
\centerline {\bf b}

\begin{center}
{\bf Table~II~(a-b).} 
{\it~The probability per plaquette of creating a handle below
(Tab.~II\,a) and above (Tab.~II\,b) the roughening temperature. 
In the first column the values of $\beta_{dual}$ appear, 
in the second the corresponding 
values of $\beta$. In the last column the values of ($\mu\times 1000$) 
are reported.}
\end {center}

Let us conclude this section by noticing that distributions of the type 
of eq.~(\ref{dist}) were already proposed both in~\cite{cgv2} and 
in~\cite{dhmmpw}. More precisely,  the following 
generalization of eq.~(\ref{dist}) was proposed:
\eq
N(g,A)=C_gA^{b(g-1)}e^{-\mu~A}
\label{general}
\en
which reduces to a Poisson distribution if $b=1$ and 
$C_g\propto 1/(g-1)!$. Both in~\cite{cgv2} and in~\cite{dhmmpw} (but for 
different reasons), values of 
$b$ different from 1 were found. In~\cite{cgv2} this was due to the fact 
that, as discussed above, the geometry of the problem was drastically 
different (and as a matter of fact the index $b$ was a function of the
lattice size $L_s$). In~\cite{dhmmpw} a value $b=1.25\pm0.01$ was found,
which could indicate the presence of a new critical behaviour near 
$\beta_c$ but, as discussed in~\cite{dhmmpw}, could also be due to 
lattice artifacts. 

In order to test the possible existence of different kinds of 
critical behaviour
we fitted the values of $N(g)$ extracted from the Montecarlo simulations 
also to eq.~(\ref{general}) keeping $b$ as a free parameter. We found 
that, in general, in a given sample, a change in  $b$ can be reabsorbed 
by a suitable change in $\mu$ still giving an acceptable agreement 
between data and theoretical distribution (even if the choice 
$b=1$ is the one that in almost all cases gives the best agreement).
However, the values of $\mu$ obtained in this way strongly depend on the 
lattice size $L_s$. For instance, for $\beta=0.3175$, if we set 
$b=1.25$ the parameter $\mu$ changes form $\mu=1.35$ to $\mu=0.80$ 
going from $L_s=14$ to $L_s=48$, thus showing that such a  choice of 
$b$ seems to be inadequate.

\subsection{ $L_s$ dependence of the mean area}
For the two values of $\beta$ and for all the lattice sizes listed in 
Table~I, we measured the mean area $A(\beta,L_s)$ (see Table~III and
Fig.~8).

$$\vbox {\offinterlineskip
\halign  { \strut#& \vrule# \tabskip=.5cm plus1cm
& \hfil#\hfil & \vrule# & \hfil# \hfil
& \vrule# & \hfil# \hfil &\vrule# \tabskip=0pt \cr \noalign {\hrule}
&& $L_s$ && $A,~\beta=0.3536$  &&  $A,~\beta=0.3175$  
 & \cr \noalign {\hrule}
&& $14$ && $411.7~(0.7)$ && $488.6~(0.8)$  & \cr \noalign {\hrule}
&& $16$ && $538.9~(1.0)$ && $639.3~(0.8)$ & 
 \cr \noalign {\hrule}
&& $18$ && $683.5~(1.2)$ && $810.1~(0.6)$ & 
 \cr \noalign {\hrule}
&& $20$ && $844.6~(1.2)$ && $1000.6~(1.2)$ & 
\cr \noalign {\hrule}
&& $22$ &&  $1022.3~(1.2)$ && $1211.2~(1.2)$ & 
\cr \noalign {\hrule}
&& $24$ && $1216.3~(1.2)$ && $1443.6~(1.2)$ &
\cr \noalign {\hrule}
&& $32$ && $$ && $2569.4~(3.4)$ &
\cr \noalign {\hrule}
&& $38$ && $$ && $3623.8~(1.6)$ &
\cr \noalign {\hrule}
&& $48$ &&  $4879.1~(4.2)$ && $5784.2~(3.6)$ &
\cr \noalign {\hrule}
}}$$

\begin{center}
{\bf Table~III.} 
{\it~Mean interface area at 
$\beta=0.3536$ and $\beta=0.3175$.}
\end {center}

We fitted these values  with a simple power law:
\eq
A(L_s)=C~(L_s)^\alpha~~.
\label{power}
\en

The  results are reported in Table~IV (see also Fig.~8):

$$\vbox {\offinterlineskip
\halign  { \strut#& \vrule# \tabskip=.5cm plus1cm
& \hfil#\hfil & \vrule# & \hfil# \hfil
& \vrule# & \hfil# \hfil &\vrule# \tabskip=0pt \cr \noalign {\hrule}
&&  && $A,~\beta=0.3536$  &&  $A,~\beta=0.3175$  
 & \cr \noalign {\hrule}
&& $\alpha$ && $2.0049~(11)$ && $2.0047~(7)$  & \cr \noalign {\hrule}
&& $C$ && $2.079~(8)$ && $2.466~(6)$ & 
 \cr \noalign {\hrule}
&& $\chi^2_r$ && $0.67$ && $0.62$ & 
 \cr \noalign {\hrule}
&& $C.L.$ && $64\%$ && $74\%$ & 
\cr \noalign {\hrule}
}}$$
 
\begin{center}
{\bf Table~IV.} 
{\it~Results of fits according to the law eq.~(\ref{power}).
In the third and the fourth rows 
the reduced $\chi^2$ and confidence level
of the fits are reported.}
\end {center}

The value of $\alpha$ is a precise indication of the two-dimensional 
nature of the interfaces that we are studying.

\subsection{ $L_s$ dependence of the mean width}

For the two values of $\beta$ and for the subset of 
underlined lattice sizes listed in Table~I, we measured the width 
$W(\beta,g,L_s)$ at fixed genus (see Figs. 9a, 9b). 
The values are given in Table~V.  
In listing our results in Table~V we only kept those samples in which at 
least 50 interfaces were present\footnote{This threshold is somewhat 
arbitrary, but we checked that exactly the same results were obtained 
setting it at 100 or 200.}.
 Hence, by looking at the empty spaces 
of Table~V, one can directly see the fact, already discussed above,
that the range of the most likely genera for an interface moves upwards 
in g
with the lattice size $L_s$ and with temperature (cf. Fig.~1). Thus, for
high temperature and/or big lattices, configurations with low genera are
unlikely (in fact configurations with 
the lowest genera were no longer observed as we moved closer to 
the critical point), while for low temperature and/or small lattices,
an interface with a high number of handles is unlikely.  
$$\vbox {\offinterlineskip
\halign  { \strut#& \vrule# \tabskip=.5cm plus1cm
& \hfil#\hfil & \vrule# & \hfil# \hfil
& \vrule# & \hfil# \hfil 
& \vrule# & \hfil# \hfil 
& \vrule# & \hfil# \hfil &
& \vrule# & \hfil# \hfil &\vrule# \tabskip=0pt \cr \noalign {\hrule}
&& $L_s$ && $g=1$  && $g=2$  && $g=3$ 
&& $g=4$ && $all~~~g$ & \cr \noalign {\hrule}
&& $14$ && $0.69~(1)$ && $0.76~(1)$ && 
$0.84~(2)$&&$$ && $0.71~(1)$ & \cr \noalign {\hrule}
&& $16$ && $0.73~(1)$ && $0.79~(1)$ && 
$0.81~(2)$&&$$ && $0.74~(1)$ & \cr \noalign {\hrule}
&& $18$ && $0.76~(1)$ && $0.80~(1)$ && 
$0.84~(2)$&&$$ && $0.77~(1)$ & \cr \noalign {\hrule}
&& $20$ && $0.79~(1)$ && $0.83~(1)$ && 
$0.87~(1)$&&$0.91~(2)$ && $0.81~(1)$ & \cr \noalign {\hrule}
&& $22$ && $0.82~(1)$ && $0.86~(1)$ && 
$0.88~(1)$&&$0.91~(3)$ && $0.84~(1)$ & \cr \noalign {\hrule}
&& $24$ && $0.84~(1)$ && $0.86~(1)$ && 
$0.89~(1)$&&$0.89~(2)$ && $0.85~(1)$ & \cr \noalign {\hrule}
&& $48$ && $0.98~(1)$ && $0.99~(1)$ && 
$0.99~(1)$&&$1.00~(1)$ && $1.00~(1)$ & \cr \noalign {\hrule}
}}$$
\vskip-0.2cm
\centerline {\bf a}

$$
\hskip-1cm
\vbox {\offinterlineskip
\halign  { \strut#& \vrule# \tabskip=.3cm
& \hfil#\hfil & \vrule# & \hfil# \hfil
& \vrule# & \hfil# \hfil 
& \vrule# & \hfil# \hfil 
& \vrule# & \hfil# \hfil 
& \vrule# & \hfil# \hfil 
& \vrule# & \hfil# \hfil &
& \vrule# & \hfil# \hfil &\vrule# \tabskip=0pt \cr \noalign {\hrule}
&& $L_s$ && $g=1$  && $g=2$  && $g=3$  && $g=4$  && $g=5$ 
&& $g=6$ && $all~~~g$ & \cr \noalign {\hrule}
&& $14$ && $1.10~(1)$ && $1.18~(1)$ &&  $1.25(1)$ && $1.34~(2)$ && 
$1.43~(4)$&& $$&& $1.15~(1)$ & \cr \noalign {\hrule}
&& $20$ && $1.22~(1)$ && $1.26~(1)$ &&  $1.31~(2)$ && $1.35~(2)$ && 
$1.37~(2)$&& $1.45~(5)$&&$1.28~(1)$ & \cr \noalign {\hrule}
&& $24$ && $1.28~(2)$ && $1.32~(2)$ &&  $1.36~(1)$ && $1.39~(2)$ && 
$1.41~(2)$&& $1.47~(3)$&&$1.35~(1)$ & \cr \noalign {\hrule}
&& $32$ && $1.38~(2)$ && $1.38~(2)$ &&  $1.41~(2)$ && $1.43~(2)$ && 
$1.45~(2)$&& $1.47~(2)$&&$1.44~(2)$ & \cr \noalign {\hrule}
&& $48$ && $$ && $$ &&  $1.52~(2)$ && $1.51~(2)$ && 
$1.54~(2)$&& $1.57~(2)$&&$1.59~(2)$ & \cr \noalign {\hrule}
}}$$
\vskip-0.2cm
\centerline {\bf b}

\begin{center}
{\bf Table~V~(a-b).} 
{\it Interface width at 
$\beta=0.3536$ (a); $\beta=0.3175$ (b). 
In the first column is reported the longitudinal lattice size $L_s$.
The following columns contain the square of the interface width $W^2$
at fixed genus. The last column contains the 
effective value of $W^2$ obtained by 
summing over all genera.}
\end {center}

We then fitted these values with the logarithmic law, eq.~(\ref{w2})
discussed above. The results of these fits are given in Table~VI.
One can see that in general all these fits have rather good confidence 
levels, but the values of $\beta_{eff}$ change remarkably with the 
genus. Only summing over all the genera one recovers the known value of 
$\beta_{eff}$~\cite{hp} 
(see last column of Table~I). Let us stress that the 
remarkable agreement of these numbers with those quoted by
M. Hasenbusch and K. Pinn~\cite{hp} (which were obtained through 
a completely independent method, namely a block spin 
renormalization approach) is an important cross-check of the reliability 
of our results. This behaviour of $\beta_{eff}$ can be 
translated into the language of interface physics by using the relation 
(\ref{kb}). It implies that, as the  genus of the interface
increases, {\it also its stiffness increases}. 
It is also interesting to notice 
that there is a natural threshold in this trend: $\beta_{eff}$ is always 
higher than the Kosterlitz-Thouless point $\beta_{eff}=2/\pi\sim 0.64$. 
What happens is again that, as the genus increases and 
$\beta_{eff}$ decreases, the probability of finding interfaces in 
thermodynamic equilibrium with the Ising Hamiltonian becomes
 smaller and smaller. This is shown by the absence of interfaces with 
genus higher than 4 and 6 respectively in Tabs. VI a and b.

$$\vbox {\offinterlineskip
\halign  { \strut#& \vrule# \tabskip=.5cm plus1cm
& \hfil#\hfil & \vrule# & \hfil# \hfil
& \vrule# & \hfil# \hfil &
& \vrule# & \hfil# \hfil &\vrule# \tabskip=0pt \cr \noalign {\hrule}
&& $Genus$ && $\beta_{eff}$  && $C_0$  && $\chi^2_r$ 
& \cr \noalign {\hrule}
&& $all~~g$ && $1.49~(6)$ && $0.09~(3)$ 
&& $1.20$ & \cr \noalign {\hrule}
&& $1$ && $1.47~(6)$ && $0.09~(3)$
&& $1.14$ & \cr \noalign {\hrule}
&& $2$ && $1.17~(6)$ && $0.27~(3)$ 
&& $0.58$ & \cr \noalign {\hrule}
&& $3$ && $0.88~(8)$ && $0.44~(4)$ 
&& $0.70$ & \cr \noalign {\hrule}
&& $4$ && $0.76~(12)$ && $0.53~(7)$ 
&& $1.17$ & \cr \noalign {\hrule}
}}$$
\vskip-0.2cm
\centerline {\bf a}

$$\vbox {\offinterlineskip
\halign  { \strut#& \vrule# \tabskip=.5cm plus1cm
& \hfil#\hfil & \vrule# & \hfil# \hfil
& \vrule# & \hfil# \hfil &
& \vrule# & \hfil# \hfil &\vrule# \tabskip=0pt \cr \noalign {\hrule}
&& $Genus$ && $\beta_{eff}$  && $C_0$  && $\chi^2_r$ 
& \cr \noalign {\hrule}
&& $all~~g$ && $2.25~(10)$ && $0.21~(5)$ 
&& $0.16$ & \cr \noalign {\hrule}
&& $1$ && $2.11~(15)$ && $0.21~(7)$
&& $0.12$ & \cr \noalign {\hrule}
&& $2$ && $1.52~(15)$ && $0.54~(7)$ 
&& $0.28$ & \cr \noalign {\hrule}
&& $3$ && $1.33~(10)$ && $0.69~(5)$ 
&& $0.41$ & \cr \noalign {\hrule}
&& $4$ && $0.90~(13)$ && $0.94~(7)$ 
&& $0.96$ & \cr \noalign {\hrule}
&& $5$ && $0.95~(17)$ && $0.94~(9)$ 
&& $2.57$ & \cr \noalign {\hrule}
&& $6$ && $0.98~(26)$ && $0.95~(15)$ 
&& $1.33$ & \cr \noalign {\hrule}
}}$$
\vskip-0.2cm
\centerline {\bf b}

\begin{center}
{\bf Table~VI~(a-b).} 
{\it $\beta_{eff}$ and $C_0$ extracted from fits to eq.~(\ref{w2}) at 
$\beta=0.3536$ (a); $\beta=0.3175$ (b). 
In the last column the reduced $\chi^2$ of the fit is reported.
In the first row the result of the fit keeping all the genera 
together is reported. In the following rows the results at fixed genus 
are reported.}
\end {center}

\section{Conclusion}
We have studied the topology of the fluid interfaces of the 3D Ising model 
in the rough phase. We found that, as the temperature increase from the 
roughening point to the Curie transition, more and more microscopic handles 
are generated which are well described by a dilute gas approximation. 
In particular,  the number of interface configurations as a 
function of the genus, $N(g)$, follows a Poisson-like distribution of
the kind $N(g)=C~\frac{{(\mu A(g))}^{g-1}}{(g-1)!}$, where 
the probability $\mu$ per unit area of creating a handle is only a 
function of the temperature, and is well fitted in 
a wide region near the roughening point by an exponentially 
decreasing function of $\beta$. The interface width $W$ follows the 
logarithmic scaling law eq.~(\ref{w2}) both separately for each genus, 
and if one sums over the genera to recover the {\it effective} 
interface. This allows one to define a genus-dependent quantity 
which behaves like the stiffness of the 
interface. An interesting feature of the stiffness defined 
in such a way is that it increases with the number of handles.

\vskip 4mm

{\bf  Acknowledgements}
We thank G. Gonnella, S. Vinti, M. Martellini, M. Spreafico and K. Yoshida
 for many helpful discussions. In particular one of us (M.C.) would like 
to thank M. Martellini and M. Spreafico for discussions on the relevance 
of the Dhen Lemma in this context. U.M. was supported in part by a
fellowship from the Italian Ministry of Foreign Affairs. She thanks the
University of Turin for hospitality during the completion of this work.

\vfill
\eject
\newpage

\centerline{\bf Figure captions}\vskip10mm
{\bf Figure~1.} {\it Genus profile  of the interfaces as a function of 
$\beta$ (note the reversed scale). 
For each value of $\beta$, 20,000 interfaces were
studied and their genus determined. For the highest temperatures
(lowest $\beta$ values)
studied, interfaces with genus up to 40 were present. For legibility,
only the curves for the first five genera are plotted, 
if the number of interfaces exceeds 100. The data were taken on a 
lattice $L_s^2 \times L_t =14^2 \times 32$.}
\vskip2mm
{\bf Figure~2.} {\it Mean area of the interfaces as a function of 
$\beta$ for the first ten genera.
Error 
bars have been drawn whenever their size exceeds the size of the symbol.}
\vskip2mm
{\bf Figure~3.} {\it Mean area difference $\Delta A$ 
between interfaces of genus $g$ and $g+1$ as a function of 
$\beta$. $\Delta A$ is essentially
independent of $g$, and is a slowly growing function of the temperature
in the rough phase (the roughening point is at $\beta \approx 0.4074(3)
$). For legibility of the graph, only scattered error bars have been 
drawn.}
\vskip2mm
{\bf Figure~4.} {\it Normalized mean area difference between interfaces 
of genus $g$ and $g+1$ in the rough phase. Below the roughening point 
($\beta \approx 0.4074$),
$\Delta A / A$ is essentially constant as a function of $\beta$. 
Scattered error bars are shown.}
\vskip2mm
{\bf Figure~5.} {\it Genus distributions of interfaces at varying values
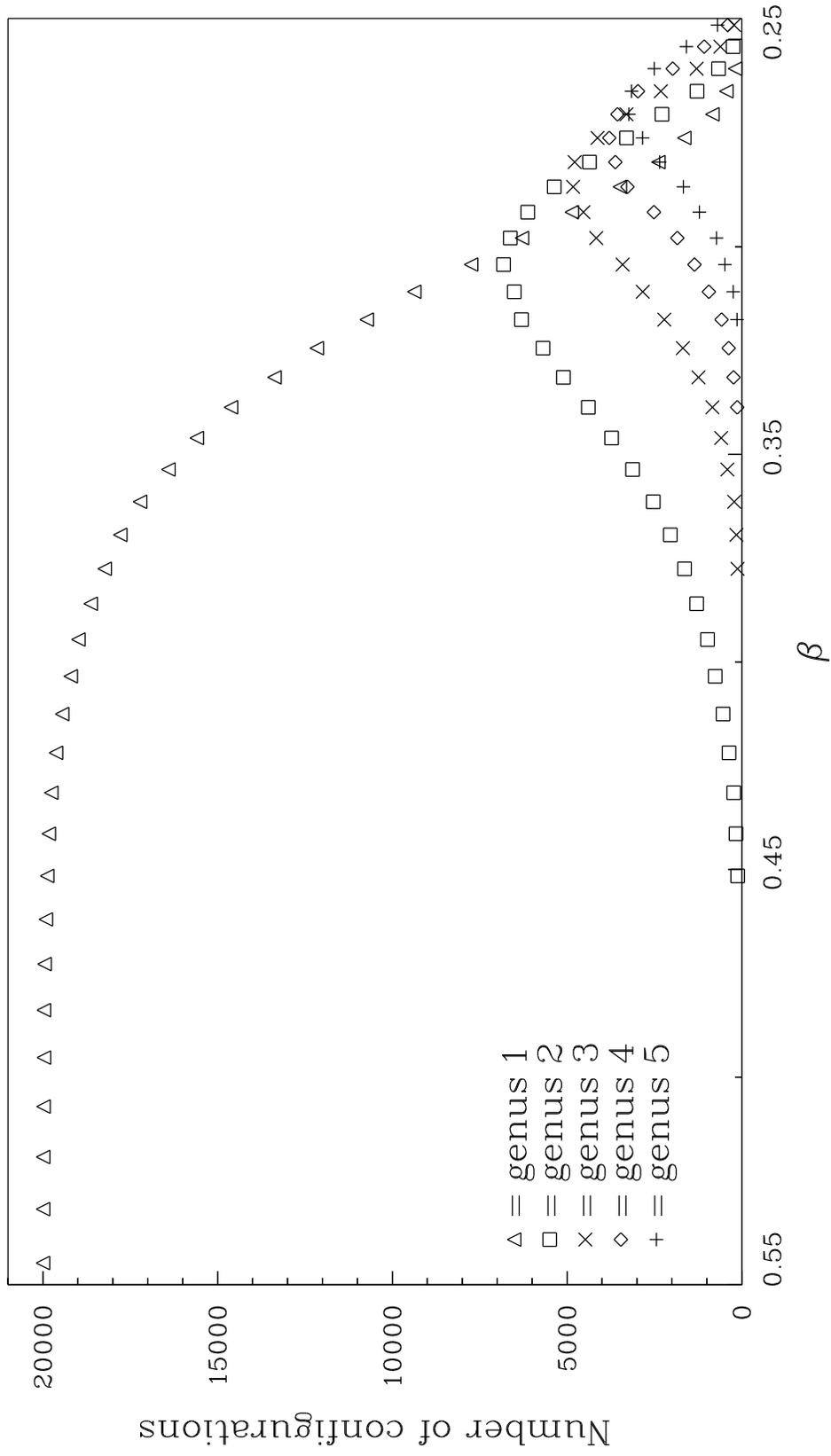
of $\beta$ and fixed lattice size. The number of configurations 
with given genus $g$,
$N(g)$ follows the Poisson-like law (16). The histograms represent the 
measured values of N(g), while the squares are the values calculated 
using eq.~(16), where the temperature-dependent parameter
$\mu$ is given in Table~II, and the measured areas were used.}
\vskip2mm
{\bf Figure~6.} {\it Same as Fig.~5, but at the fixed inverse temperature
$\beta = 0.3175$, and with varying longitudinal 
lattice size $L_s$. The parameter $\mu$ is always the same.}
\vskip2mm
{\bf Figure~7.} {\it An illustration of the law (18) for $\mu(\beta)$,
the probability per plaquette of having a handle. This law was based on 
the simplest possible assumption concerning the dependence of $\mu$
on $\beta$. The line is eq.~(18) with $l_h \simeq 14$ and $\mu_0
\simeq 7.5$, and the squares are the values for $log(\mu)$
extracted from (16) and the measured values for $N(g)$ (number 
of interfaces)
and $A(g)$ (area of interfaces) at fixed lattice size as a 
function of $\beta$. 
Due to the lack of statistics for higher genera, we do not quote
errors for $\mu$.}
\vskip2mm
{\bf Figure~8.} {\it Mean area $A$ of interfaces as a 
function of lattice 
size $L_s$ and inverse temperature $\beta$. On the horizontal axis 
we have $log(L_s)$, and on the vertical axis $log(A)$. The straight 
lines represent the power law fits (20) (see Table~IV). The errors are
microscopic on the scale of the figure. 
The measured 
points are for the effective interface (summed over all genera).}
\vskip2mm
{\bf Figure~9a-b.} {\it Squared interface width 
$W^2$ versus $log(L_s)$ at
$\beta = 0.3536$ (a) and $\beta = 0.3175$. 
At $\beta = 0.3536$, statistics for the first 
four genera were available, and for $\beta = 0.3175$, for the first six.
The measured values for the interface width 
(defined in (15)) are plotted, with fits according to eq.~(7) 
corresponding to the dashed lines. The measured squared width of 
the effective interface, summed over all genera, is also shown
together with the fit (the whole-drawn line).
Here one gets a clear hint as to how the
various topologies "sum up" to give an effective interface.}  
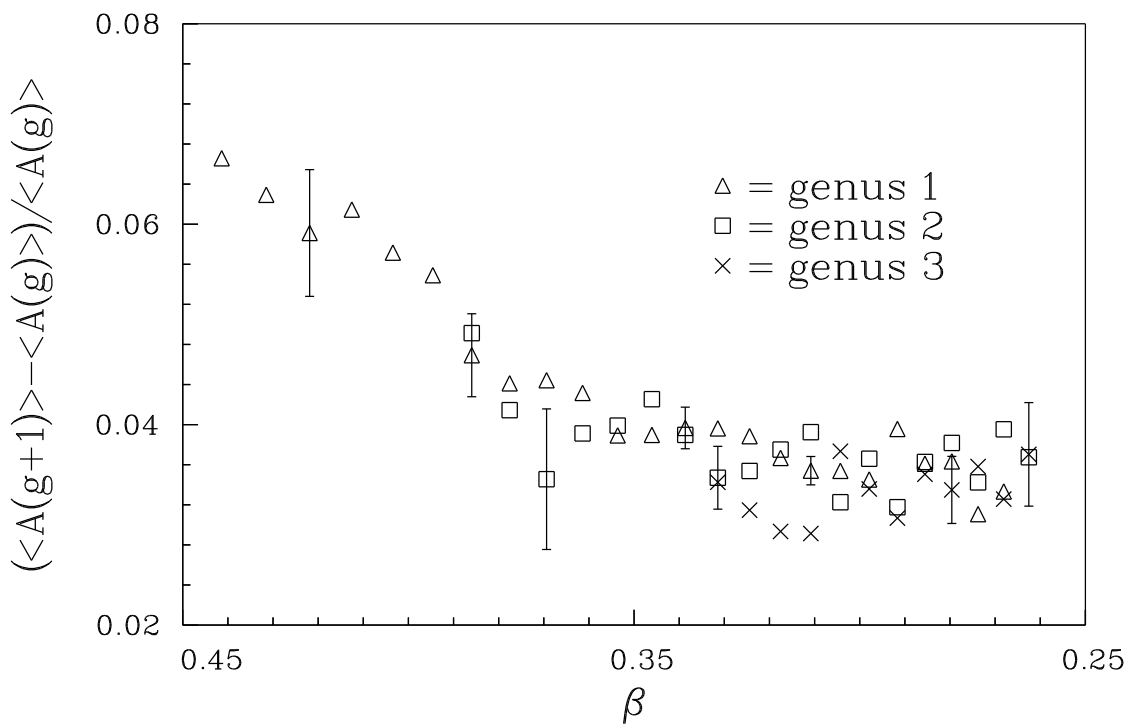
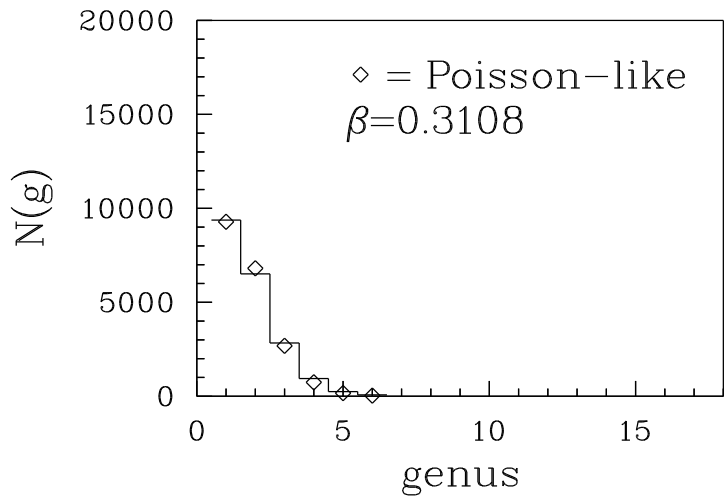
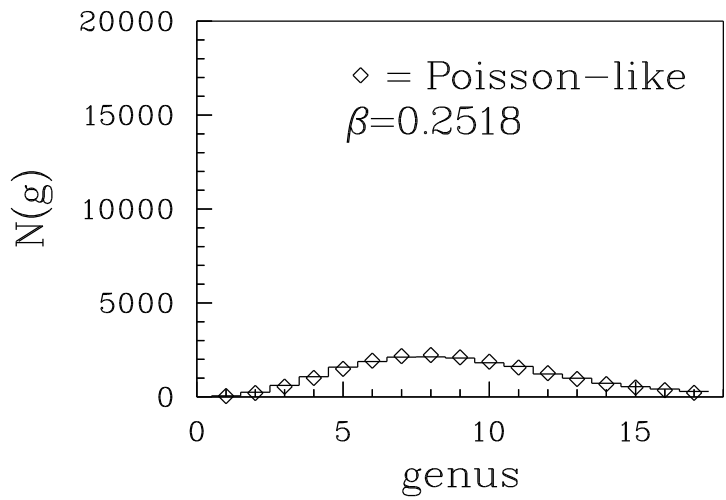
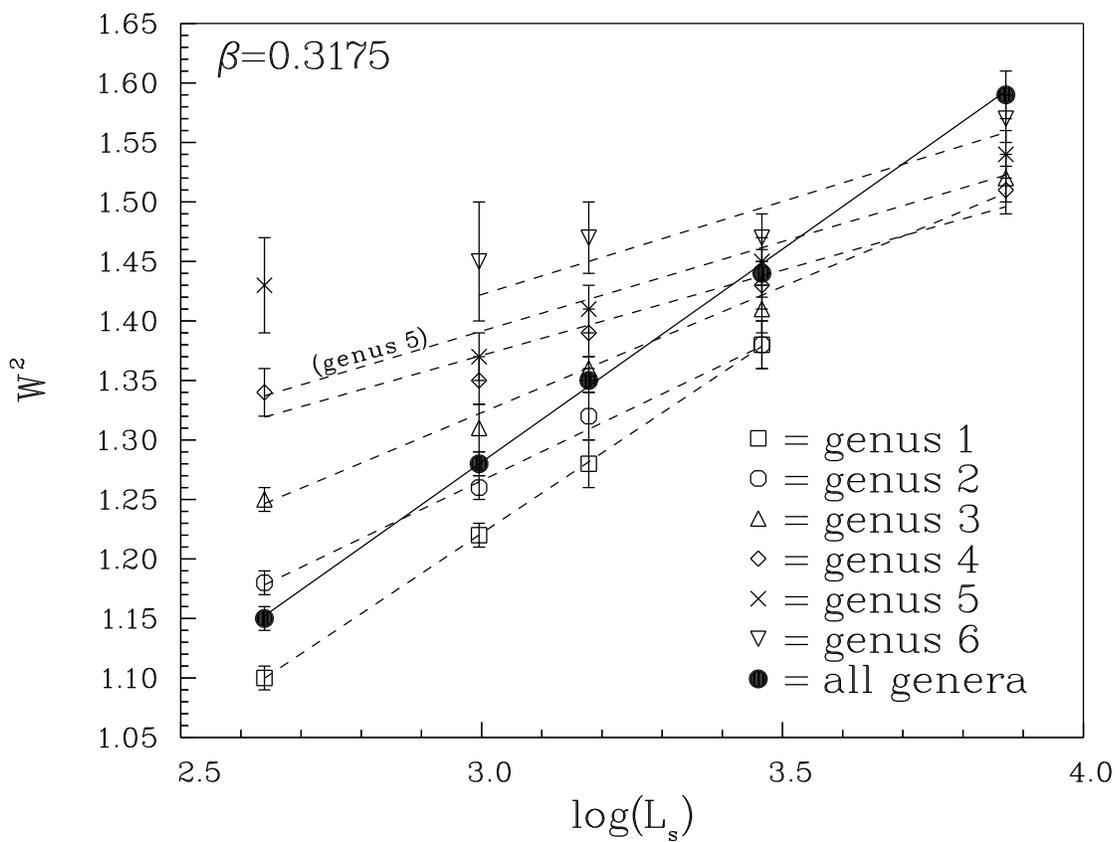
\end{document}